# Micrometer-thick, atomically random $Si_{0.06}Ge_{0.90}Sn_{0.04}$ for silicon-integrated infrared optoelectronics

S. Assali,[1,a)] A. Attiaoui,[1] S. Koelling,[1] M. R. M. Atalla,[1] A. Kumar,[1] J. Nicolas,[1] Faqrul A. Chowdhury,[1] C. Lemieux-Leduc,[1] and O. Moutanabbir[1,a)]

[1] Department of Engineering Physics, École Polytechnique de Montréal, C. P. 6079, Succ. Centre-Ville, Montréal, Québec H3C 3A7, Canada

[a)] Authors to whom correspondence should be addressed: simone.assali@polymtl.ca, oussama.moutanabbir@polymtl.ca

## ABSTRACT

A true monolithic infrared photonics platform is within reach if strain and bandgap energy can be independently engineered in SiGeSn semiconductors. Herein, we investigate the structural and optoelectronic properties of a 1.5 μm-thick $Si_{0.06}Ge_{0.90}Sn_{0.04}$ layer that is nearly lattice-matched to a Ge on Si substrate. Atomic-level studies demonstrate high crystalline quality and uniform composition and show no sign of short-range ordering and clusters. Room temperature spectroscopic ellipsometry and transmission measurements show direct bandgap absorption at 0.83 eV and a reduced indirect bandgap absorption at lower energies. $Si_{0.06}Ge_{0.90}Sn_{0.04}$ photoconductive devices operating at room temperature exhibit dark current and spectral responsivity (1 A/W below 1.5 μm wavelengths) similar to Ge on Si devices, with the advantage of a near-infrared band gap tunable by alloy composition. These results underline the relevance of SiGeSn semiconductors in implementing a group IV material platform for silicon-integrated infrared optoelectronics.

# I. INTRODUCTION

SiGeSn ternary semiconductors are silicon-compatible materials providing the capability to independently engineer the bandgap energy and lattice parameter in a similar fashion to the well-established III-V and II-VI semiconductors.[1–5] The availability of these Sn-containing group IV semiconductors (SiGeSn) would enable key building blocks for silicon-integrated monolithic photonic and optoelectronic devices.[6] With this perspective, major efforts have been recently expended to develop SiGeSn materials. The vast majority of these studies have been focusing on GeSn binary semiconductors leading to the demonstration of a range of devices operating at the short-wave infrared (SWIR: 1.4-3 μm) and mid-wave IR (MWIR: 3-8 μm) wavelengths.[5,7–14] Direct bandgap absorption and emission are now at reach using GeSn grown on Si at Sn contents above ~10 at.%, which is more than 10 times higher than the ~1 at.% equilibrium solubility of Sn in Ge.[15] These achievements are clear evidence that the thermodynamic constraints inherent to GeSn can be overcome yielding device quality material by controlling nonequilibrium growth kinetics in strain-engineered multi-layer heterostructures. By minimizing the epitaxial compressive strain and keeping the growth temperature well below 400 °C, phase separation and segregation of Sn are avoided and high-quality GeSn epilayers are obtained.[16–18]

In ternary SiGeSn semiconductors, the thermodynamic constraints are exacerbated as the equilibrium solubility of Sn in Si is about one order of magnitude lower than in Ge.[19] By increasing Si incorporation, the bandgap of the material increases and the optical properties can be tuned to shorter wavelengths in the near-IR (NIR: 0.75-1.4 μm).[1,5,6] At the same time, the possibility to control the lattice parameter by varying the Si/Sn ratio allows to decouple strain and composition effects, which is critical to implement lattice-matched SiGeSn/GeSn heterostructures. A simultaneous increase in the Si and Sn content can be precisely engineered to preserve the lattice-matched growth on a substrate, while progressively increasing the SiGeSn band gap.[5] Earlier reports demonstrated the monolithic growth of SiGeSn at thicknesses below 600 nm and bandgap energies in the 0.8-1.2 eV range.[1,3,20–25] Increasing the thickness of SiGeSn layers above 1 μm while preserving their optoelectronic properties is crucial to fully exploit the attractive properties of these emerging semiconductors.[6,26] It is important to note that the growth of thicker layers must also preserve the composition uniformity and material quality. In fact, it has been shown that Sn incorporation is greatly affected by strain relaxation as the layers grow thicker,[17] leading to a compositional gradient. Circumventing these challenges and achieving thick and uniform layers is of paramount importance not only to enhance the IR absorption, but also to implement hybrid III-V/SiGeSn multijunction solar cells[27,28] and improve the performance of GeSn-based light emitting devices.[4,13,14,29–34] For instance, in recently demonstrated electrically-pumped $Ge_{0.89}Sn_{0.11}$ lasers, the $Si_{0.03}Ge_{0.89}Sn_{0.08}$ barriers are limited to a thickness of 250 nm.[14] Thicker SiGeSn barriers would reduce scattering of the lasing modes, while a higher Si content would increase carrier confinement in the GeSn active layer at room-temperature, further enhancing radiative emission and device efficiency.[6] A type-I band alignment between $Si_{0.03}Ge_{0.89}Sn_{0.08}/Ge_{0.89}Sn_{0.11}$ with a band offset higher than 100 meV was achieved in this GeSn-based LED heterostructure.[14] Moreover, despite SiGeSn being an indirect band gap material at near-infrared wavelengths, it has a lower Γ- to L-valley energy separation as compared to Ge.[5] This is expected to improve direct bandgap absorption and thus enhance the performance of IR photodetectors, while providing low leakage current that could enable the operation under high bias voltages. In general, the monolithic growth of Sn-containing group IV semiconductors is achieved using Ge virtual substrate on Si (Ge-VS/Si).

Si is a smaller atom than Ge and Sn, and its incorporation in the lattice results in a higher compressive strain that could compromise the metastable state of the Sn-saturated alloys.[17,35]

Herein, we investigate the structural and optoelectronic properties of a 1.5 μm-thick $Si_{0.06}Ge_{0.90}Sn_{0.04}$ layer that is nearly lattice-matched to the Ge-VS/Si wafer. High crystalline quality, uniform composition, and the absence of Sn segregation and clustering are demonstrated at the-atomic level in $Si_{0.06}Ge_{0.90}Sn_{0.04}$, where a n-type carrier concentration of $-5\times10^{16}$ $cm^{-3}$ is estimated. Spectroscopic ellipsometry and transmission measurements at room temperature indicate direct bandgap absorption at 0.83 eV, with a reduced indirect bandgap absorption at lower energies. $Si_{0.06}Ge_{0.90}Sn_{0.04}$ photoconductive devices exhibit identical dark current to that measured for Ge devices, which is 10 times lower than similar devices made of lattice-mismatched $Ge_{1-x}Sn_x$ ($x\geq10$ at.%) heterostructures. The relatively low dark current in $Si_{0.06}Ge_{0.90}Sn_{0.04}$ is relevant to high bias voltage applications with reduced noise and power dissipation and highlights the benefit of a lattice-matched growth on device performance. Lastly, a spectral responsivity of 1.0-1.1 A/W in the 1.0-1.5 μm wavelength range was recorded at room temperature in $Si_{0.06}Ge_{0.90}Sn_{0.04}$, slightly above the value obtained for similar Ge devices.

# II. EXPERIMENTAL DETAILS

Samples were grown on 4-inch Si (100) wafers in a low-pressure chemical vapor deposition (CVD) reactor using ultra-pure $H_2$ carrier gas, and 10 % monogermane ($GeH_4$), 2 % disilane ($Si_2H_6$) and tin-tetrachloride ($SnCl_4$) precursors. First, a 1.9 µm-thick Ge-VS was grown with a two-temperature step process (450/600 °C) and post-growth thermal cyclic annealing (>800 °C). Next, a ~10 nm-thick $Ge_{0.99}Sn_{0.01}$ nucleation layer was grown at 360 °C for 10 minutes using a Ge/Sn ratio (in gas phase) of ~2500.[17] $Si_2H_6$ was then introduced in the chamber (Si/Sn ratio in gas phase of ~15), without modifying the $GeH_4$ and $SnCl_4$ supply, for the 6 hours growth of the 1.5 µm-thick $Si_{0.06}Ge_{0.90}Sn_{0.04}$ layer. Reference Ge-VS samples with thicknesses of 1.9 µm and 3.0 µm were also grown. 20µm×20µm atomic force microscopy (AFM) maps were acquired to estimate the surface roughness of the samples. Atom Probe Tomography (APT) measurements were performed in a LEAP 5000XS from Cameca using a 355 nm laser producing picosecond pulses at a variable repetition rate of a few 100 kHz. The tip-shaped samples for APT were prepared using a FIB based procedure introduced in Ref.[36] in a FEI Helios FIB using a Gallium ion beam with energies between 30 and 5 kV. Spectroscopic ellipsometry measurements were performed using a J. A. Woollam® RC2-XI ellipsometer with a dual rotating compensator. Data were acquired from 250 to 2500 nm with 1 nm step and at incident angles ranging from of 60° to 85° with a 5° step. Near-Brewster angle (between 70° and 80°) SE measurements were undertaken to build the optical model that is based on a three-layer system ($Si_{0.06}Ge_{0.90}Sn_{0.04}$, Ge-VS, Si wafer) and includes a surface roughness layer. For each sample, a second set of measurements was taken at different orientation to verify the in-plane isotropy of the film. The samples were confirmed to be isotropic in the *x-y* plane. Polarized transmission and reflectance measurements (for *s* and *p* polarizations, at the incident angle of 75°) were also performed on bare substrates first, and then on substrates with as-grown Ge-VS using a Cary 50 UV-VIS universal measurement spectrophotometer (UMS) at identical angle of incidence as the SE. A thorough description of the measurement and modelling strategy of the optical properties of the semiconductors is presented in the Supplementary Material. Polarized Transmission Spectroscopy was performed using a double beam spectrophotometer equipped with rotating sample and detector stages (Cary 7000 UMS from Agilent Technologies). The reflectance (R) and transmittance (T) data were acquired at various angles of incidence without moving the sample (from 60 ° to 85 ° so that R and T matches the SE AOI). Polarization-dependent measurements were limited to the 250-2500 nm range, which matches the SE spectral range. The unpolarized transmission data are fitted simultaneously with the SE parameters to validate the built optical model. Back-to-back C-V devices were fabricated with electron-beam deposition of $SiO_2$/Ti/Au (20/10/70 nm) and characterized using a Keithley 4200a parameter analyzer connected to a RT probe station. The fabrication of metal-semiconductor-metal photoconductive devices consisted of a single photolithography layer to deposit Ti/ Au (10/70 nm) metal contacts using an electron-beam evaporation tool. The I-V measurements were acquired using the aforementioned parameter analyzer. The photocurrent was measured at 1.55 μm wavelength showing a calibrated responsivity of 0.57, 0.66, 0.95 and 1.02 for as-grown $Si_{0.06}Ge_{0.90}Sn_{0.04}$, Ge (1.9 μm), Ge (3.0 μm), and annealed $Si_{0.06}Ge_{0.90}Sn_{0.04}$, respectively, at 25 V. The spectral responsivity was measured using a Bruker Vertex 80 FTIR spectrometer equipped with a calibrated NIR light source, connected to a lock-in amplifier operating at 500 Hz (chopper along the NIR light optical path).

# III. RESULTS AND DISCUSSION

## III.a) Structural properties down to the atomic-level.

The cross-sectional scanning transmission electron microscopy (STEM) image for the 1.5 μm-thick $Si_{0.06}Ge_{0.90}Sn_{0.04}$ layer grown on a 1.9 μm-thick Ge-VS is shown in Figure 1a. The parameters obtained from the structural analysis are listed in Table 1. We note that the simultaneous incorporation of Si and Sn yields a nearly lattice-matched growth on Ge, with a larger bandgap in $Si_{0.06}Ge_{0.90}Sn_{0.04}$ than Ge, as discussed below. In low-resolution TEM (Figure 1b), few misfit dislocations are detected at the $Si_{0.06}Ge_{0.90}Sn_{0.04}$-Ge interface, and no threading dislocations are observed in $Si_{0.06}Ge_{0.90}Sn_{0.04}$ at the TEM imaging scale. The Ge-Si interface exhibits higher dislocation density (Figure 1c), which is expected for a lattice-mismatched growth. The high crystalline quality of the $Si_{0.06}Ge_{0.90}Sn_{0.04}$/Ge heterostructure is demonstrated by X-ray diffraction spectroscopy (XRD). Figure 1d displays a typical 2θ-ω scan performed around the (004) XRD order. The Si and Ge peaks are visible at 69.10 ° and 66.06 °, respectively, while the $Si_{0.06}Ge_{0.90}Sn_{0.04}$ peak is recorded at 65.72 °. In a strain-free GeSn layer, a peak shift to lower angles corresponds to an increase in Sn content in the layer, and as Si is incorporated the diffraction peak shifts to larger angles. Considering that both $Si_{0.06}Ge_{0.90}Sn_{0.04}$ and Ge layers have a comparable thickness, similar XRD peak intensities and full width at half maxima (0.029 ° and 0.027 °, respectively) indicate that the crystalline quality of $Si_{0.06}Ge_{0.90}Sn_{0.04}$ is almost identical to that of Ge. To evaluate the residual strain in the as-grown layers, reciprocal space mapping (RSM) measurements around the asymmetrical (224) XRD peak were performed, as exhibited in Figure 1e. Si-Ge and Ge-Sn bowing parameters of 0.026 Å [37] and 0.41 Å [7] were considered, respectively. The small in-plane tensile strain $\varepsilon_{||} \leq 0.2$ % in the Ge originates from the differences between the thermal expansion coefficients of Si and Ge. The $Si_{0.06}Ge_{0.90}Sn_{0.04}$ layer has an in-plane lattice constant of $a_{||} = 5.669$ Å that is almost identical to the Ge-VS, while the $a_{\perp} = 5.680$ Å, is slightly larger than the one of the Ge-VS (5.651 Å). This results in a very low compressive strain $\varepsilon_{||} = -0.2$ %. The small broadening of the $Si_{0.06}Ge_{0.90}Sn_{0.04}$ peak and the absence of interference fringes suggest that a limited plastic relaxation occurs in the system, most likely triggered by the presence of threading dislocations originating from the Ge-VS. Overall, the TEM (Fig. 1b) and XRD results indicate that $Si_{0.06}Ge_{0.90}Sn_{0.04}$ is nearly lattice-matched to the Ge-VS. A Si:Sn ratio of 1.5 is estimated for the 1.5 µm-thick $Si_{0.06}Ge_{0.90}Sn_{0.04}$ layer. The measured 1.5 ratio falls within the 0.5-4.5 range previously reported for SiGeSn layers that are lattice-matched on bulk Ge and on Ge-VS.[1–4,22,26]

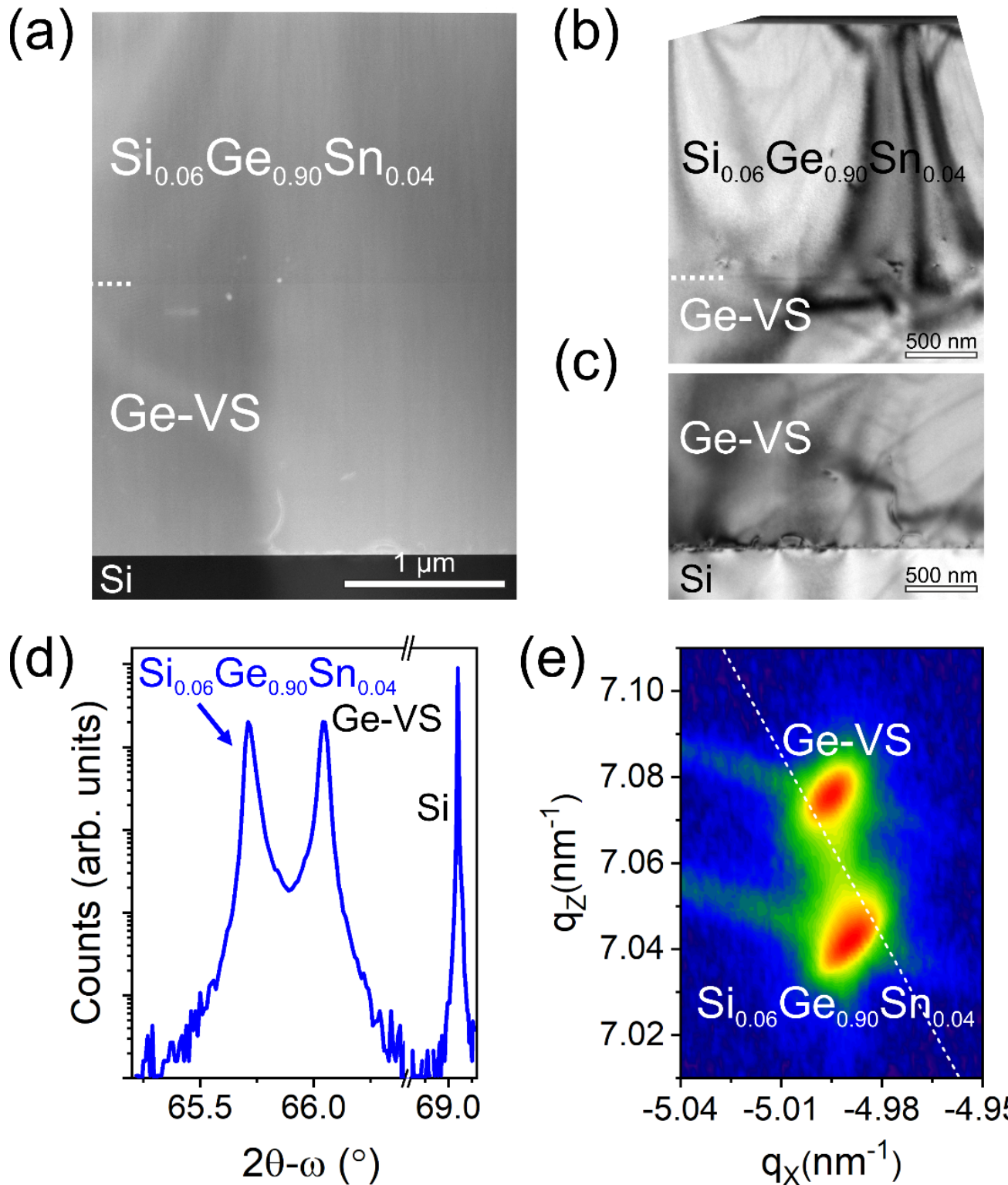

**Figure 1.** (a) STEM image of the 1.5/1.9 µm-thick $Si_{0.06}Ge_{0.90}Sn_{0.04}$/Ge-VS heterostructure grown on a Si wafer. (b-c) Low-resolution TEM images acquired across the heterostructure showing the $Si_{0.06}Ge_{0.90}Sn_{0.04}$/Ge-VS (b) and Ge-VS/Si (c) interfaces. (d) 2θ-ω scan around the (004) X-ray diffraction order. (e) RSM around the asymmetrical (224) reflection.

The atomic-level chemical composition of the grown layers was studied by atom probe tomography (APT). The 3D reconstruction of a typical APT tip is shown in Figure 2a, where different atoms are color-coded: Si (green), Ge (blue), and Sn (red). In the APT compositional profile (Figure 2b), the uniform incorporation of Si and Sn is confirmed at a measured composition of 4 at.% and 6 at.% across the top ~1400 nm-thick layer, respectively. The measured content fluctuations are less than 0.3 at.%. Note, that the lower signal-to-noise ratio close to the sample surface originates from the reduced volume in the upper portion of the APT tip. A compositional gradient is visible in the 150 nm-thick region near the SiGeSn/Ge interface (Fig. 2c), resulting from strain-driven incorporation and diffusion. Based on the APT 3D maps, statistical analyses in Figure 3 were performed on the entire $Si_{0.06}Ge_{0.90}Sn_{0.04}$ layer to investigate the relative positions of Si, Ge, and Sn atoms and assess whether a random alloy is formed as it is typically assumed in theoretical modeling of the optical and electronic properties.[5,26] First, the nearest-neighbor (NN) distributions are evaluated up to the 10$^{th}$ order (solid spheres) and compared with a randomly shuffled assignment of the element to the same underlying point cloud of atom positions reconstructed from the measured data (dashed black curves).[38] The results of the NN analysis are shown in Figure 3a for the Si-Si (top) and Sn-Sn (bottom) pair of atoms. The agreement between experimental data and randomized data proves the absence of Sn and Si clusters containing more than 10 atoms,[39,40] and the absence of precipitates as expected for a locally random alloy. The APT data show a compositional gradient lower than 0.2 at.%/μm for both Si and Sn atoms, thus confirming the absence of significant concentration gradients across a ~1400 nm-thick portion of the $Si_{0.06}Ge_{0.90}Sn_{0.04}$ layer. Frequency distribution analysis (FDA) (Fig. 3b) comparing the measured distribution of atoms with a randomly shuffled arrangement, is also in agreement with a locally random alloy.[41]

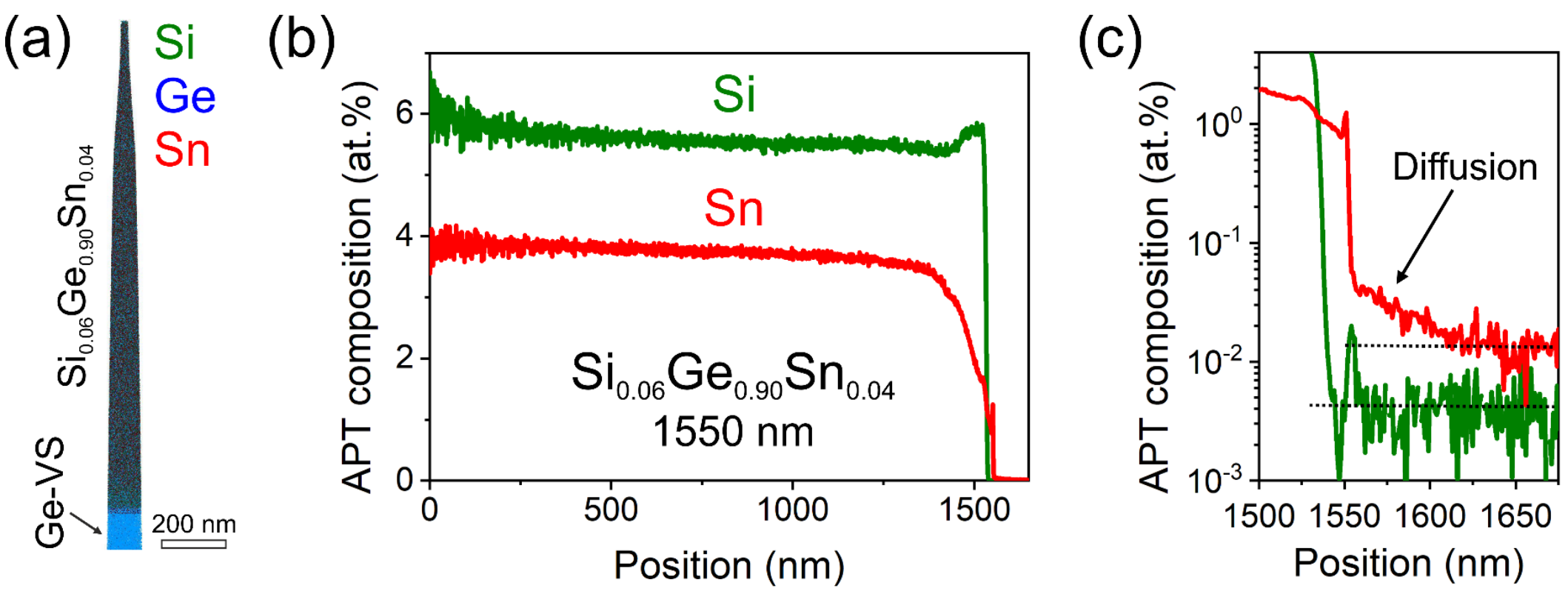

**Figure 2.** (a-b) APT 3D reconstruction (a) and compositional profiles for the Si, Sn atoms (b). (c) Enlarged view of (d) showing the interface region between $Si_{0.06}Ge_{0.90}Sn_{0.04}$ and Ge-VS.

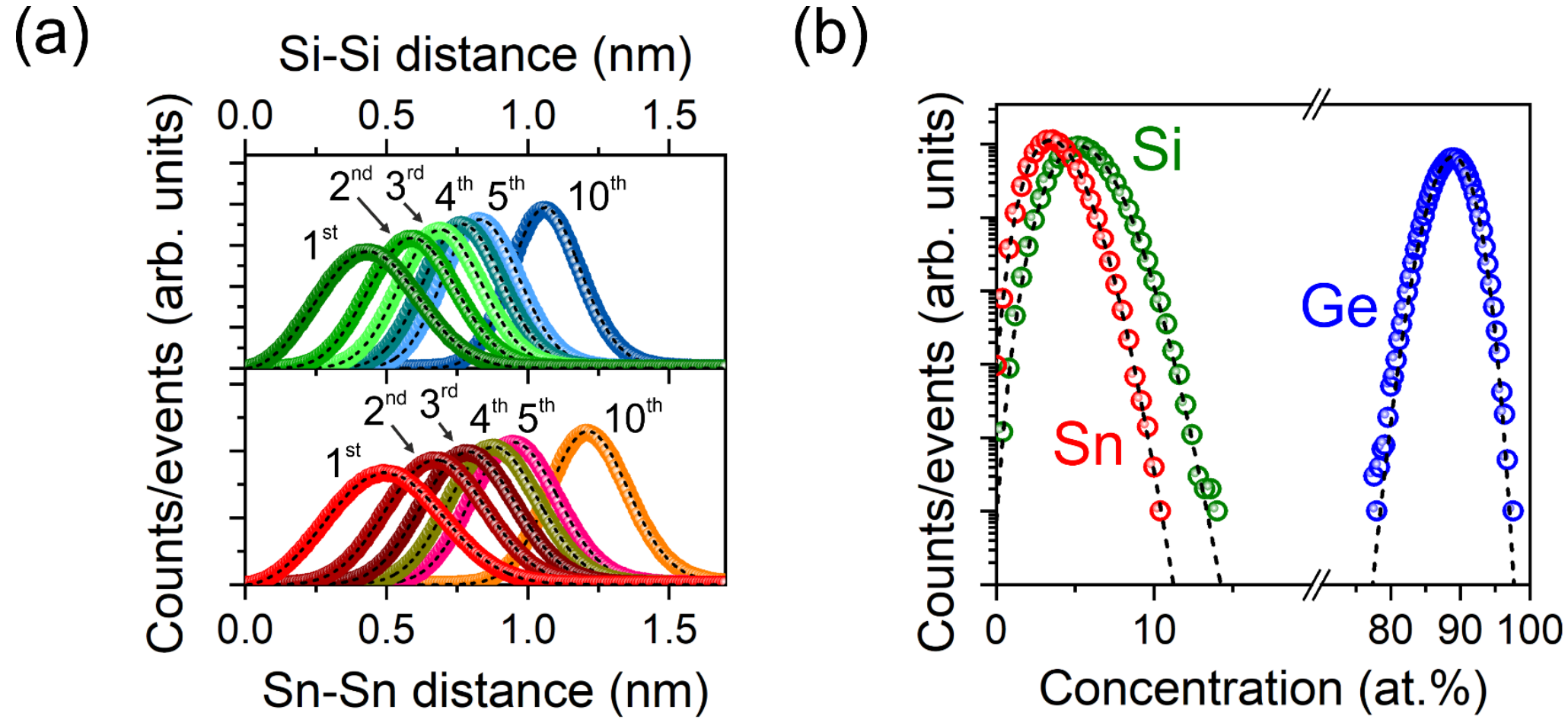

**Figure 3.** (a) NN distributions evaluated up to the $10^{th}$ order (solid spheres) for the Si-Si (top curves) and Sn-Sn (bottom curves) pairs. The theoretical curves for a perfect random alloy are shown as dashed black curves. (b) FDA analysis for 250 atoms. The binomial distributions expected for a random alloy are shown using dashed curves.

### III.b) Analysis of residual free carrier density.

The demonstrated high structural quality of $Si_{0.06}Ge_{0.90}Sn_{0.04}$ paves the way to investigate the potential of this material system in optoelectronic devices. As a first step, a more detailed knowledge of the type of conductivity and free carrier concentration in the epilayer would provide valuable information for device processing and optimization. A low carrier concentration, eventually tunable to reach compensation, would reduce the dark current, enhance the response time in photodetector devices, and enable high-speed operation by reducing parasitic capacitance. To estimate the residual free carrier density, capacitance-voltage (C-V) measurements were performed in a back-to-back geometry (Figure 4a). The obtained results are shown in Figure 4b. P-type carrier concentration in the $2\cdot10^{16}$-$2\cdot10^{17}$ $cm^{-3}$ range was measured for Ge epilayers grown at different thicknesses ($7\cdot10^{16}$ $cm^{-3}$ from Ref.[42] at a thickness of 1.6 μm is shown in Figure 4b). No effect of the thickness on the carrier concentration in Ge was observed. Interestingly, n-type carrier concentration of $-5\pm3\cdot10^{16}$ $cm^{-3}$ was measured in $Si_{0.06}Ge_{0.90}Sn_{0.04}$. This behavior is consistent and always observed in all SiGeSn samples regardless of their thickness (not shown). Moreover, frequency-dependence C-V measurements show switching characteristics that are similar to state-of-the-art n-type metal oxide semiconductor capacitors (MOSCaps), which further supports the n-type conductivity obtained in our SiGeSn sample (Supplementary Material S1). This behavior differs from the p-type conductivity recorded for SiGeSn layers at a Si (Sn) content up to 10 at.% (14 at.%) and a thickness up to 600 nm.[25] While only one experimental study is available for SiGeSn so far, it is also worth mentioning that GeSn layers show consistently a p-type carrier concentration in the $1$-$5\cdot10^{17}$ $cm^{-3}$ range.[25,42,43]

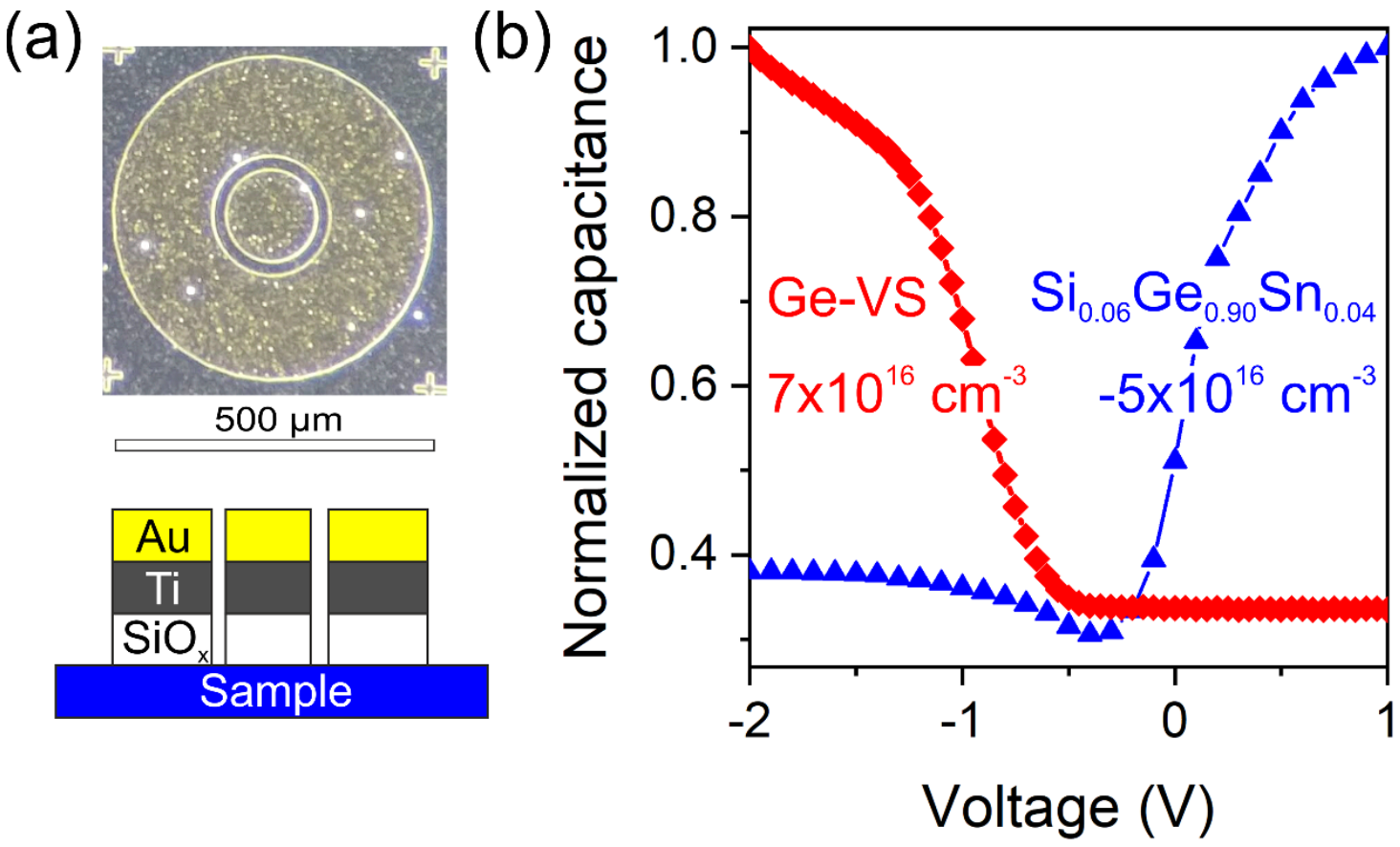

**Figure 4.** (a) Optical micrograph and schematics of the back-to-back metal-oxide-semiconductor device. (b) C-V curves for $Si_{0.06}Ge_{0.90}Sn_{0.04}$ and Ge-VS acquired at a frequency of 1 MHz. The latter curve was adapted from Ref.[42].

In a broad sense, due to the relatively low growth temperature of metastable Sn-containing group IV semiconductors, point defects, such as divacancies and vacancy-clusters,[44,45] can be electrically active and their nature can impact the conductivity type and residual free carrier concentration. However, point defects in SiGeSn and their effects on the electronic behavior are yet to be investigated. Moreover, as intrinsic GeSn shows consistently a p-type behavior, the incorporation of Si should be examined as a potential factor that may influence the background doping. This can possibly occur by affecting the defect complexes and their charge states. It is, however, important to note that p-type residual doping is observed in epitaxial Ge and does not change in undoped SiGe semiconductors at Si contents in the 10-30 at.% range, where the carrier concentration lies below $5 \cdot 10^{15}$ $cm^{-3}$.[46] Therefore, it is even more challenging to evaluate the role of Si in SiGeSn, where the lattice properties are very different. Background doping in the CVD reactor can also be a plausible element to investigate. Herein, we would like to stress that our undoped GeSn[7,42] layers and SiGeSn were grown in the same reactor and they were found to exhibit an opposite conductivity. Additionally, it is noteworthy that the SiGeSn samples are kept at 360 °C for a few hours during growth, hence at ≥50 °C higher temperature compared to GeSn.[7,42] This extended thermal annealing process, combined with a low growth rate (~4 nm/min), could play a significant role in diffusion and annihilation of point defects in SiGeSn. In-depth studies are needed to provide the necessary theoretical framework to elucidate the origin of the residual free carriers and their type in SiGeSn. Experimental investigations of vacancies and vacancy-like defects using positron lifetime annihilation spectroscopy combined with electrical characterization would be a good start to shed light on this phenomenon.[44,45]

## III.c) Optical properties.

Spectroscopic ellipsometry (SE) and UV-VIS spectrophotometry measurements were combined to characterize the as-grown layers, as shown in Figure 5, and the most relevant critical point values obtained from the analysis are listed in Table 1. Based on these two techniques, an independent and consistent evaluation of the direct bandgap is obtained (see Supplementary Material S2 for more details). The real and imaginary parts $\varepsilon_1$ and $\varepsilon_2$ of the complex dielectric function are displayed in Figure 5a. The $E_0$, $E_1$, $E_1+\Delta_1$, and $E_2$ critical points (CPs) in the joint density of states are identified in the curves. Because of the higher surface roughness (RMS) in the $Si_{0.06}Ge_{0.90}Sn_{0.04}$ (up to 30 nm) layer compared to the Ge-VS (<5 nm), a reduction in the $E_2$ peak is observed. The direct band gap of the samples is estimated from the $E_0$ CP transition by fitting the square of the absorption coefficient $\alpha$ using a Tauc plot procedure, as shown in Figure 4b.[47,48] Direct bandgap values of 0.836±0.002 eV, 0.779±0.008 eV, and 0.785±0.002 eV are obtained for $Si_{0.06}Ge_{0.90}Sn_{0.04}$, 1.9 µm-thick Ge and 3.0 µm-thick Ge samples, respectively. The small tensile strain ($\varepsilon_{||} \leq 0.20$ %) in the Ge epilayer results in a direct bandgap energy being lower than that of bulk Ge (0.80 eV), where the energy difference with the latter decreases with increasing epilayer thickness (strain decreases from 0.20 % to 0.15 %). A steep cut-off at lower energies is visible in $Si_{0.06}Ge_{0.90}Sn_{0.04}$, with a reduced contribution from the Urbach tail compared to Ge. This indicates a suppression of the indirect bandgap absorption in favor of the direct one (inset in Figure 5b, logarithmic scale for $\alpha$). Following a similar approach to that described in Refs.[10,49], the Urbach tail, which is attributed to the transition between bandtail states, defined as electronic states right above the valence band and the conduction band, was quantified. To that end, the absorption coefficient was fitted, below the band gap, to the following relationship:

$$\alpha = A\sqrt{\Delta E/2}\,\exp\left(\left(h\nu - E_g^{\Gamma}\right)/\Delta E\right) \quad , \text{ (Eq. 1)}$$

where $A$ and $\Delta E$ (Urbach energy) are material-dependant parameters and $E_g^{\Gamma}$ is the direct bandgap. $\Delta E$ values of 5.7±0.3 meV, 12.2±0.5 meV, and 14.8±0.5 meV were obtained for the $Si_{0.06}Ge_{0.90}Sn_{0.04}$, 1.9 µm-thick Ge, and 3.0 µm-thick Ge layers, respectively. Our estimated Urbach energy for Ge is in excellent agreement with the work of Tran *et al.*[10] It is noteworthy that $\Delta E$ is reduced by half in $Si_{0.06}Ge_{0.90}Sn_{0.04}$ compared to Ge, thus confirming the high crystalline quality of the ternary alloy. This observation is similar to previous reports in GeSn with compositions up to 17 at.%[9,10] and combined with the reduced indirect bandgap absorption it highlights the excellent optical quality of $Si_{0.06}Ge_{0.90}Sn_{0.04}$ enabled by the lattice-matched epitaxy.

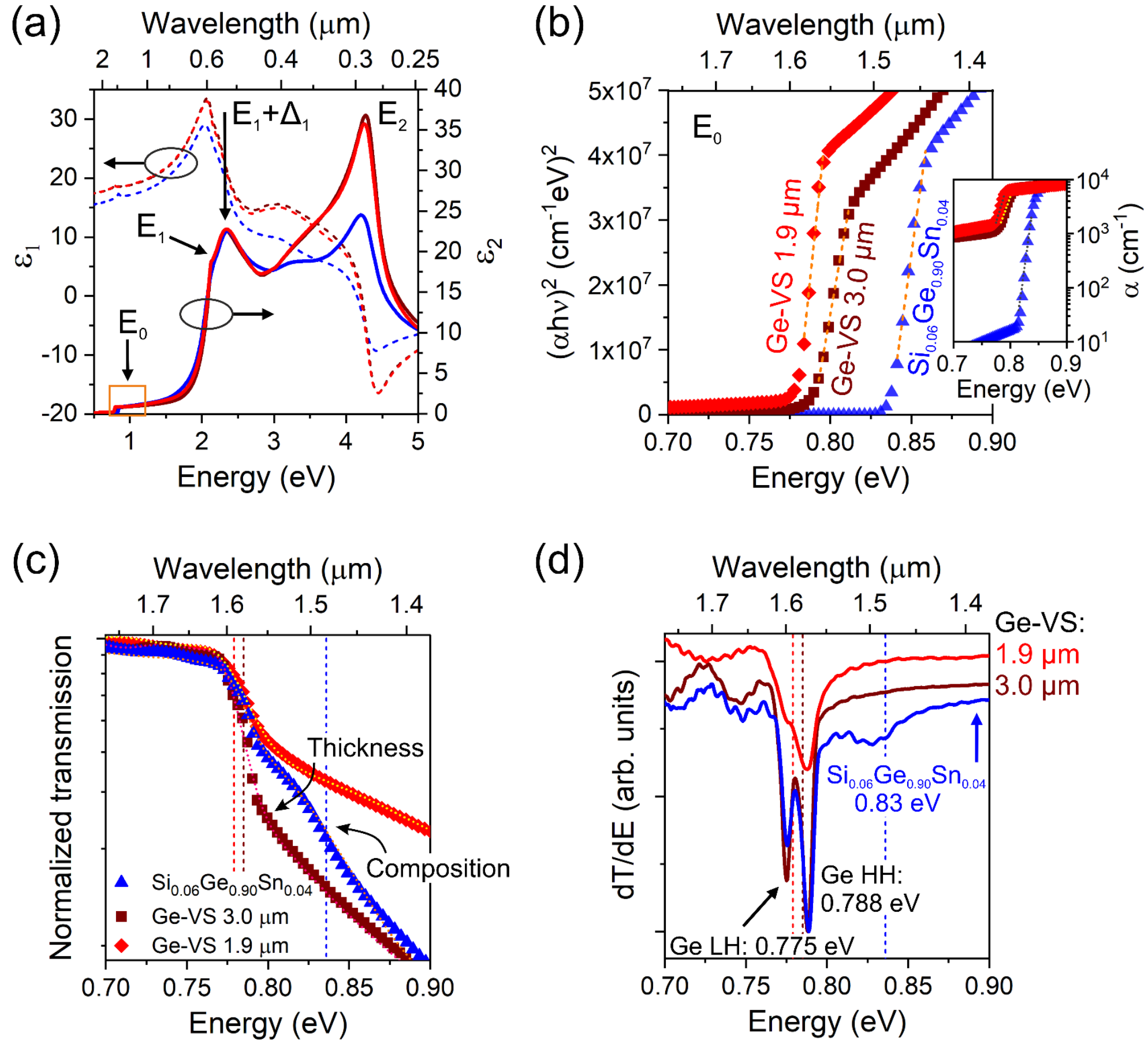

**Figure 5.** (a) Complex dielectric function ($\varepsilon_1$ (dashed line) and $\varepsilon_2$ (solid line)) of 1.5/1.9 μm-thick $Si_{0.06}Ge_{0.90}Sn_{0.04}$/Ge-VS (blue curves), 1.9 μm-thick Ge-VS (red curves), and 3.0 μm-thick Ge-VS (dark red curves) samples. (b) Tauc plot $(\alpha h\nu)^2$ and linear fit for the bandgap (dashed lines). Inset: Urbach tail fit (dotted lines) for α. (c) Transmission measurements (solid curves) and simulated curves calculated from the SE built optical model (dotted curves). (d) Derivative of the T as a function of energy showing independent HH and LH bands. The vertical dashed lines in (c-d) are the estimated $E_0$ values from SE.

The SE-estimated band gaps are compared (vertical dashed line in Figure 5c) with those from transmission spectra, as displayed in Figure 5c. While absorption in Ge takes place at 0.77 eV, the $Si_{0.06}Ge_{0.90}Sn_{0.04}$ contributes to additional absorption above 0.83 eV. The bandgap is quantified through fitting the transmission spectra. The dashed red lines in Figure 5c highlight the result of the fitting process, confirming the accuracy of the optical model (Figure 5a). A transfer matrix method (TMM) was undertaken to theoretically quantify the transmission.[50] The obtained values agree with the peaks observed in the first derivative of the transmission ($dT/dE$), as displayed in Figure 5d. The two distinct peaks at 0.775±0.005 and 0.788±0.005 eV in the 3.0 µm-thick Ge-VS (less resolved in the thinner sample) correspond to the Γ-transitions associated with light-holes (LH) and heavy-holes (HH), respectively. Because of the residual tensile strain, the LH-HH degeneracy is lifted and the bandgap energy is slightly reduced relative to bulk Ge.[51] We note that the presence of Fabry-Perot interference fringes in the transmission measurements, with spacing dependent on the thickness, does not affect the estimation of the 0.83 eV direct bandgap of $Si_{0.06}Ge_{0.90}Sn_{0.04}$, which closely matches the value obtained from the SE analysis.

## III.d) Room-temperature spectral photoresponsivity.

To investigate the electrical properties of $Si_{0.06}Ge_{0.90}Sn_{0.04}$, current-voltage (I-V) curves are acquired on photoconductive devices (PD) in Figure 6a. The optical micrograph of the $Si_{0.06}Ge_{0.90}Sn_{0.04}$ PD and related dark current as a function of the applied bias are displayed in Figure 6a. A Schottky behavior is recorded in the as-grown device (blue curve) with a knee voltage of ~6 V. By performing rapid thermal annealing (RTA) at 380 °C for 30 s an Ohmic behavior is obtained (green curve), which is most likely due to Fermi level unpinning as a result of metal diffusion into $Si_{0.06}Ge_{0.90}Sn_{0.04}$. Figure 6b compares the I-V curves of the annealed $Si_{0.06}Ge_{0.90}Sn_{0.04}$ PD with the 1.9 µm-thick (dotted dark red curve) and 3.0 µm-thick (dashed red curve) Ge PDs. Similar dark current is measured in these three sets of devices. At first, this finding indicates that the lattice-matched $Si_{0.06}Ge_{0.90}Sn_{0.04}$ growth on Ge-VS (Figure 1) limits growth defects in the heterostructure, thereby preventing any additional defect-related leakage current. It is worth noting that growth defects have already been demonstrated to increase dark and leakage currents in Ge devices,[52] while no similar studies are available for SiGeSn to date. Nevertheless, the similarities between SiGeSn and Ge devices were not expected. Indeed, considering that Ge layers were grown at a higher temperature of 600 °C (*vs*. 360 °C for $Si_{0.06}Ge_{0.90}Sn_{0.04}$) and were subjected to thermal annealing above 800 °C, a lower amount of point defects is expected in Ge compared to SiGeSn. However, similar experimental values (but opposite sign) for the intrinsic carrier concentration $n_i$ are obtained for both materials (Figure 3). This seems to indicate that a possible difference in concentration of point defects in these materials does not dramatically affect the absolute value of $n_i$. When compared to lattice-mismatched GeSn PDs at Sn contents above 10 at.% grown in the same CVD reactor,[42] the dark current and intrinsic carrier concentration in Ge and $Si_{0.06}Ge_{0.90}Sn_{0.04}$ devices are one order of magnitude lower. Multiple factors can lead to a higher intrinsic carrier concentration $n_i$ in GeSn, such as the smaller bandgap (<0.5 eV), a change in the density of states, and possibly a higher density of point defects. All these factors, combined with a higher density of dislocations in the lattice-matched multi-layer heterostructure, result in a higher leakage current in GeSn compared to $Si_{0.06}Ge_{0.90}Sn_{0.04}$ or Ge.

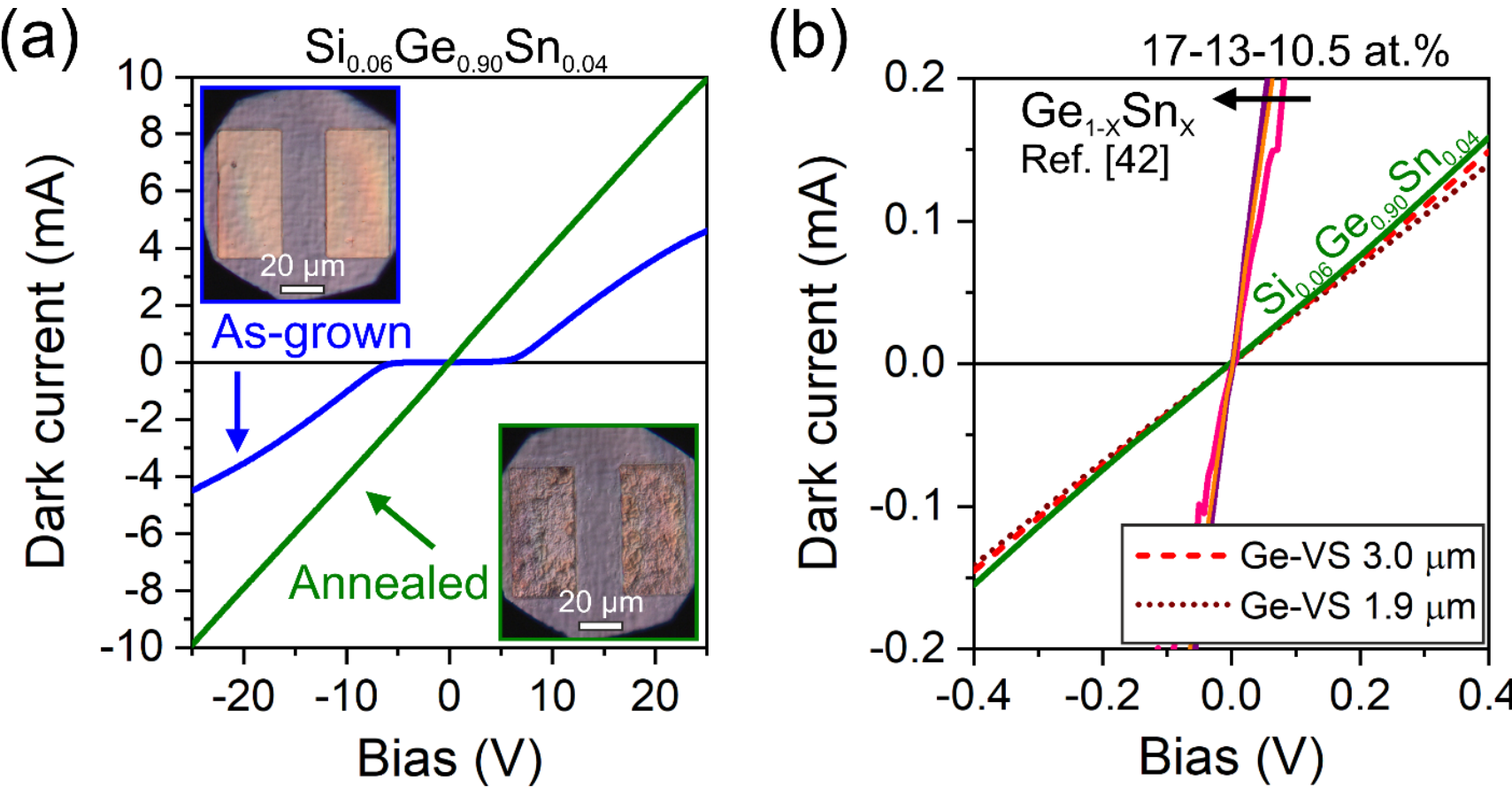

**Figure 6.** (a) Optical micrographs and I-V curves for as-grown (blue) and annealed (green) $Si_{0.06}Ge_{0.90}Sn_{0.04}$ PD devices. (c) I-V curves for annealed (green) $Si_{0.06}Ge_{0.90}Sn_{0.04}$, 1.9 µm-thick (dashed red) and 3.0 µm-thick (dotted dark red) Ge-VS PDs. The curves for $Ge_{1-x}Sn_x$ (x=10.5-13-17 at.%, pink-orange-purple curves) PDs from Ref.[42] are shown as a comparison.

Room-temperature spectral responsivity curves of $Si_{0.06}Ge_{0.90}Sn_{0.04}$ and Ge PDs are shown in Figure 7. A bias voltage of 25 V (effective electric field 15 $kV/cm^{-1}$) was selected for the photocurrent measurements to show the potential for high voltage applications based on $Si_{0.06}Ge_{0.90}Sn_{0.04}$, such as single-photon avalanche photodetectors.[53] However, operation at lower voltages is also possible in annealed devices. We first focus on the Ge PDs, where a steep increase in responsivity above 0.75 eV and a rather flat curve in the 0.82-1.15 eV range are observed, followed by a small decrease at higher energies (Figure 7a and 7b upper panel). Above bandgap carrier thermalization[54] could reduce PD efficiency at the highest detection energies. By increasing the Ge thickness from 1.9 to 3.0 µm, the responsivity increases across the whole measurement range, with a peak responsivity of ~0.96 A/W at 0.83 eV (*i.e.* 1.5 µm wavelength) in the 3.0 µm-thick Ge PD. The increase in responsivity with layer thickness indicates that the larger amount of photogenerated carriers in a thicker Ge epilayer can be efficiently collected by the PD on top, thus showing limited carrier trapping at misfit and threading dislocations in the heterostructure. The higher responsivity at larger thickness is also in agreement with the reduction in threading dislocation density as Ge thickness increases.[55] The derivative of the responsivity as a function of the energy is shown in Figure 7b (bottom panel). In Ge PD, a single peak at 0.78 eV is obtained, hence at the same energy of the LH-HH doublet peak in Figure 5d.

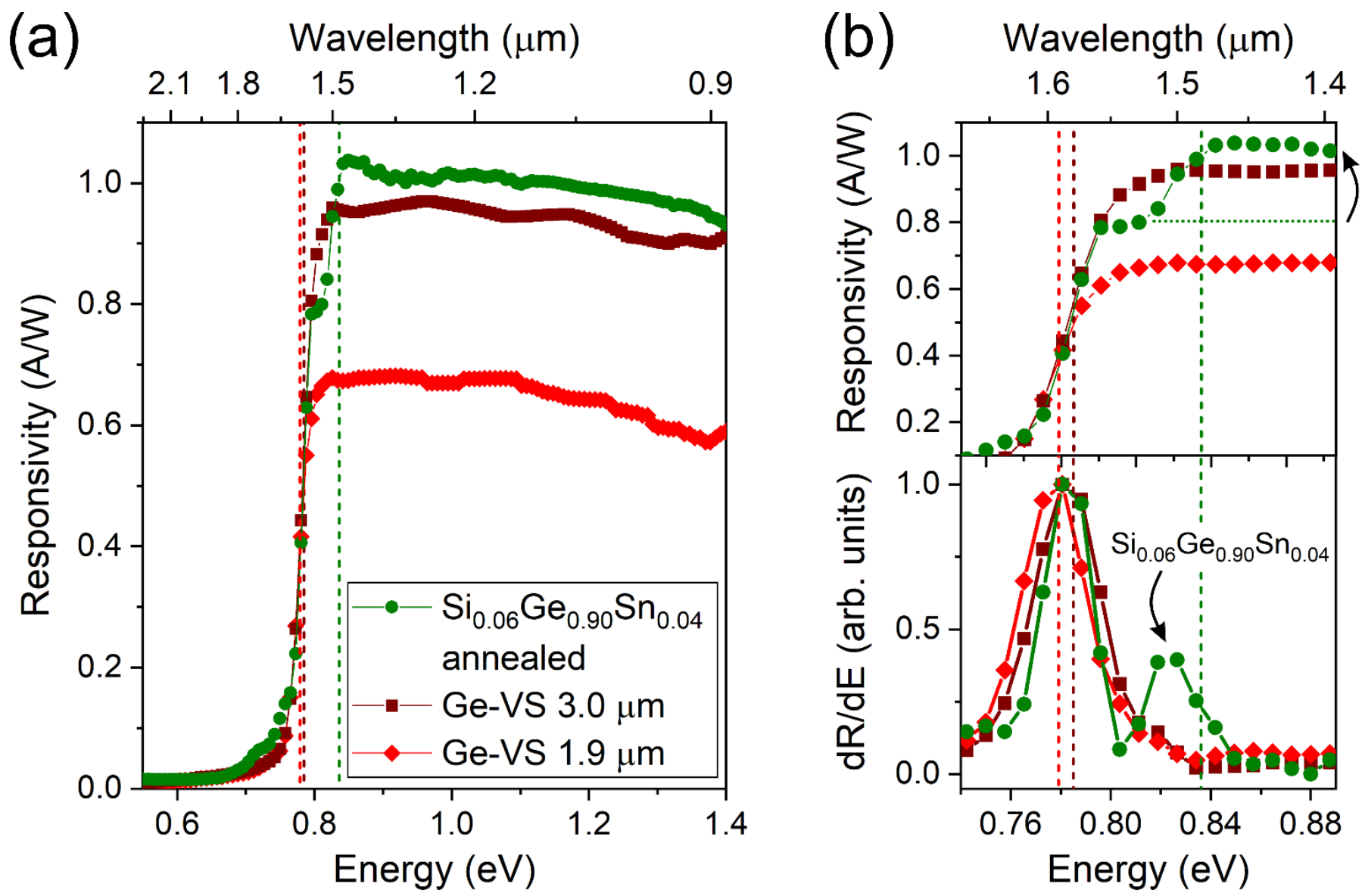

**Figure 7.** (a) Responsivity curves for all samples acquired at a bias voltage of 25 V. (b) Enlarged view of (a) and corresponding derivative as a function of the energy. Vertical dashed lines indicate the $E_g$ values estimated from the optical measurements in Figure 5.

In the as-grown $Si_{0.06}Ge_{0.90}Sn_{0.04}$ PD device, a peak responsivity of 0.57 A/W is measured at 0.84 eV (not shown). Interestingly, this value almost doubles up to 1.04 A/W after RTA, thus with a similar enhancement ratio to what is observed in the I-V curves (Figure 6a). This shows that the Schottky barrier at the contacts has been diminished as a result of the annealing process. As expected, an additional $Si_{0.06}Ge_{0.90}Sn_{0.04}$ absorption edge is visible at 0.82 eV, which is ~40 meV above the Ge-related peak. This energy shift is in perfect agreement with SE and transmission measurements (Figure 3) and it is unaffected by post-growth RTA. Increasing the Si content would increase the energy shift toward a larger bandgap, which offers high potential for photovoltaic applications.[27,56] The fact that the spectral responsivity of annealed $Si_{0.06}Ge_{0.90}Sn_{0.04}$ PDs is comparable to the Ge-based ones is rather surprising given the thermal budget applied to each set of these devices. Indeed, as mentioned above, Ge layers were subjected to post-growth thermal cyclic annealing in $H_2$ above 800 °C to reduce threading dislocation density and improve crystallinity, which should increase PD performance. In $Si_{0.06}Ge_{0.90}Sn_{0.04}$, the rapid thermal annealing was limited to 380 °C (30 s) to maintain the crystalline quality of these metastable layers.[57] Nevertheless, optimized thermal annealing protocols for SiGeSn could further increase device efficiency.

| | Thickness (nm) | Strain (%) | $a_0$ (Å) | $E_0$ [SE] (eV) | $E_1$ [SE] (eV) | $E_1 + \Delta_1$ [SE] (eV) | $\Delta E$ [SE] (meV) | $E_\Gamma$ [T] (eV) | $E_\Gamma$ [PD] (eV) |
|---|---|---|---|---|---|---|---|---|---|
| $Si_{0.06}Ge_{0.90}Sn_{0.04}$ | 1550<br>(±20) | -0.13<br>(±0.02) | 5.669 (∥)<br>5.680 (⊥)<br>(±0.001) | 0.836<br>(±0.002) | 2.099<br>(±0.03) | 2.301<br>(±0.03) | 5.7<br>(±0.3) | 0.829<br>(±0.02) | 0.823<br>(±0.04) |
| Ge-VS | 3000<br>(±20) | 0.15<br>(±0.02) | 5.666 (∥)<br>5.651 (⊥)<br>(±0.001) | 0.785<br>(±0.002) | 2.121<br>(±0.01) | 2.305<br>(±0.01) | 14.8<br>(±0.5) | 0.775 (HH)<br>0.788 (LH)<br>(±0.01) | 0.784<br>(±0.04) |
| Ge-VS | 1900<br>(±20) | 0.20<br>(±0.02) | 5.669 (∥)<br>5.650 (⊥)<br>(±0.001) | 0.779<br>(±0.008) | 2.115<br>(±0.01) | 2.298<br>(±0.02) | 12.2<br>(±0.5) | 0.775 (HH)<br>0.788 (LH)<br>(±0.01) | 0.777<br>(±0.04) |

**Table 1.** List of the structural and optoelectronic parameters obtained from TEM, APT, SE, transmission (T), and photodetector (PD) responsivity measurements.

The results above demonstrate that high quality, micrometer-thick $Si_{0.06}Ge_{0.90}Sn_{0.04}$ semiconductors can deliver similar performance as Ge on Si devices with similar background doping levels ($10^{16}$ $cm^{-3}$ range). While the results presented in this work are mainly focused on $Si_{0.06}Ge_{0.90}Sn_{0.04}$, similar optoelectronic properties are expected at higher Si and Sn contents, provided that a uniform composition is achieved through the epilayer. The possibility to independently engineer the lattice parameter and bandgap energy, paves the way for SiGeSn as a suitable material for IR detection, with a strong potential for the future optimization of group IV integrated optoelectronic devices. Unlike conventional II-VI and III-V semiconductors used in IR applications, SiGeSn can be tuned from indirect to direct bandgap material by controlling the Sn/Si composition ratio, thus enabling advanced carrier lifetime engineering at a tunable (or fixed) energy of the optical absorption.[58] Moreover, the SiGeSn directness ($\Delta = E_L - E_\Gamma$) increases with Sn content leading to higher absorption, while preserving the long carrier lifetime of an indirect bandgap semiconductor. This would lead to electrons to scatter from direct to indirect valleys in the conduction band, thus suppressing radiative and Auger recombination and potentially resulting in a quasi-direct semiconductor regime with $k$-space charge separation.[6,59] High carrier collection efficiency is required for the development and integration of a 1.0 eV SiGeSn junction in hybrid Ge/InGaAs/InGaP multijunction solar cells as an alternative approach to conventional III-V/Ge devices.[27,56] In the nearly lattice-matched $Si_{0.06}Ge_{0.90}Sn_{0.04}$/Ge heterostructure, optically-generated carriers in Ge can diffuse through the $Si_{0.06}Ge_{0.90}Sn_{0.04}$ heterostructure due to the low dislocation density, and be collected on the sample surface. This is a promising feature that can be beneficial for the integration in scalable solar cells.

## IV. CONCLUSION

In summary, we unveiled the structural and optoelectronic properties of a 1.5 μm-thick $Si_{0.06}Ge_{0.90}Sn_{0.04}$ layer grown on a Ge-VS/Si wafer. High crystalline quality, uniform composition, and the absence of short-range ordering effects and clusters were demonstrated for the $Si_{0.06}Ge_{0.90}Sn_{0.04}$ epilayer. Room temperature direct bandgap absorption at 0.83 eV in $Si_{0.06}Ge_{0.90}Sn_{0.04}$ was investigated by combining spectroscopic ellipsometry, transmission, and photocurrent measurements. The optical analysis shows a reduction of the indirect bandgap absorption at lower energies and of the Urbach energy. Similar dark current was estimated in $Si_{0.06}Ge_{0.90}Sn_{0.04}$ and Ge photoconductive devices and found to be one order of magnitude lower than that of binary GeSn multi-layer devices. The low dark current in $Si_{0.06}Ge_{0.90}Sn_{0.04}$ further highlights the benefit of a lattice-matched growth on device performance, thus paving the way for high bias voltage applications with reduced noise and power dissipation. Lastly, spectral responsivity of 1.0-1.1 A/W in the 0.82-1.15 eV range (*i.e.,* 1.5-1.0 μm wavelength range) was recorded at room temperature in $Si_{0.06}Ge_{0.90}Sn_{0.04}$, hence comparable to similar devices made of Ge. These results, combined with the possibility to further increase the Si, Sn content and tune the band gap at shorter wavelengths, cement the relevance of SiGeSn alloys as versatile building blocks for silicon-integrated infrared optoelectronics and multi-junction solar cells on a Si wafer.

## SUPPLEMENTARY MATERIAL

See the supplementary materials for additional details on the frequency-dependent C-V measurements and for the spectroscopic ellipsometry measurements protocol.

## ACKNOWLEDGEMENTS

The authors thank J. Bouchard for the technical support with the CVD system, B. Baloukas for support with the SE measurement. O.M. acknowledges support from NSERC Canada (Discovery, SPG, and CRD Grants), Canada Research Chairs, Canada Foundation for Innovation, Mitacs, PRIMA Québec, and Defence Canada (Innovation for Defence Excellence and Security, IDEaS).

## CONFLICT OF INTEREST

The authors declare no conflict of interest.

## DATA AVAILABILITY

The data that support the findings of this study are available from the corresponding author upon reasonable request.

# Micrometer-thick, atomically random $Si_{0.06}Ge_{0.90}Sn_{0.04}$ for silicon-integrated infrared optoelectronics

## Supplementary Material

S. Assali,[1,*] A. Attiaoui,[1] S. Koelling,[1] M. R. M. Atalla,[1] A. Kumar,[1] J. Nicolas,[1] Faqrul A. Chowdhury,[1] C. Lemieux-Leduc,[1] and O. Moutanabbir[1,*]

[1] Department of Engineering Physics, École Polytechnique de Montréal, C. P. 6079, Succ. Centre-Ville, Montréal, Québec H3C 3A7, Canada

## S1. Frequency-dependent C-V measurements.

At a fixed frequency of 1 MHz (Fig. 3b) the active carrier concentration in the samples was estimated according to the following equation:

$$N_{sub} = \frac{2}{q\varepsilon_s A^2\left(\frac{\Delta 1/c^2}{\Delta V_g}\right)}, \quad \text{(Eq. 1)}$$

where $\varepsilon_s$ is the relative permittivity of the semiconductor, A is the area of the gate, and $\left(\frac{\Delta 1/c^2}{\Delta V_g}\right)$ is the averaged slope of the linear part in transition region of the C-V curves. The relative permittivity of GeSn was assumed to be similar to pure Ge with a value of $\varepsilon_s = 16$. We note that a correction factor of 0.8 was used for the area in the B2B devices based on the dimensions of the ring geometry of the capacitor.

The representative multifrequency measurement on SiGeSn (back-to-back, B2B) metal oxide semiconductor capacitors (MOSCaps) in Figure S1 reveal well-formed, steep C-V characteristics without noticeable overall frequency dispersion and gate leakage. By varying the frequency over three orders of magnitude, a single minimum is observed at -0.9 V, which indicates that only the device of interest (with smaller area) is dominant in the SiGeSn B2B MOSCap device. Although a slight frequency dispersion in accumulation region can be observed, which might be attributed to series resistance, the device shows switching characteristics that are similar to the state-of-the-art n-type MOSCaps. This further supports the estimation of the n-type doping in the material. The MOSCaps from low bandgap semiconductors typically depict bumps in their response curves which are attributed to the presence of defect states in the vicinity of the semiconductor/oxide interface ($D_{it}$). The SiGeSn MOSCaps not only show U-shaped low frequency behavior at low to moderate frequencies (>1 kHz) at room temperature (ascribed to enhanced minority carrier generation, originating from the small bandgap at dark condition), but its response is also free from any weak inversion bump (related to moderate $D_{it}$) or bump in the depletion region (related to significantly higher $D_{it}$).

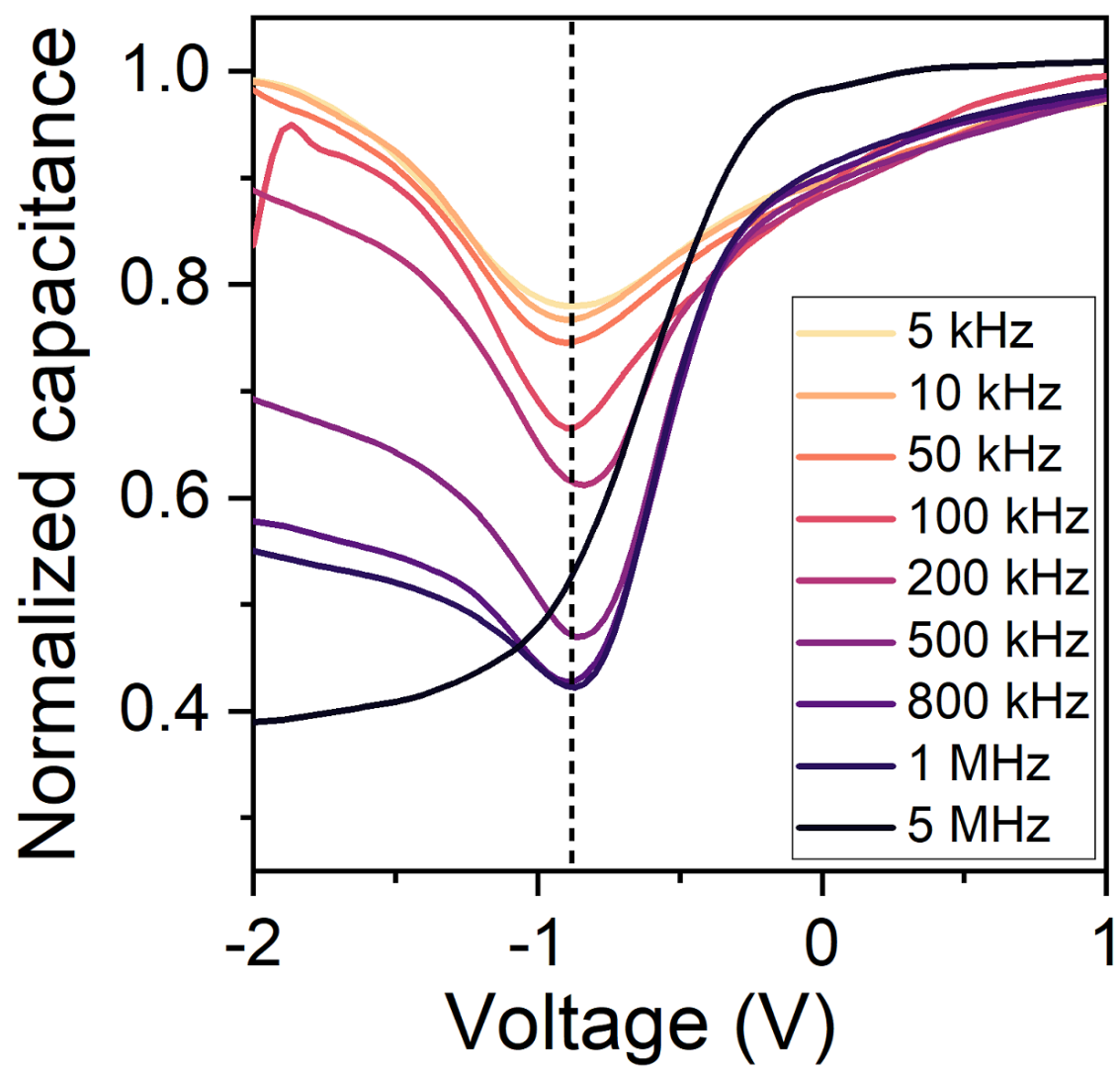

**Figure S1:** Multifrequency C-V measurement on SiGeSn.

## S2. Spectroscopic ellipsometry measurements protocol.

In spectroscopic ellipsometry (SE) measurements, a beam of linearly polarized light is emitted onto the surface of a solid sample at an oblique angle of incidence (AOI). SE measure the amplitude ratio and the phase difference between $p$- and $s$-polarizations, from which the optical constants can be deduced. The ellipsometric angle $(\Psi, \Delta)$ were acquired from 250 to 2500 nm with 1 nm step at 6 AOI (between 60° and 85° with a 5° step), on a J.A. Woollam© RC2-XI spectroscopic ellipsometer, with a dual rotating compensator. The dielectric functions of the SiGeSn samples on Ge were obtained by modeling the ellipsometry data and doing a multi-sample analysis (MSA). The dielectric functions for the $GeO_2$[1] and $SnO_2$[2] were used in tabulated form from published data whereas the dielectric function of Ge was measured independently with SE and modelled from two as-grown CVD-samples with 1.9 µm and 3 µm layers thicknesses. Tensile strain in both Ge layers was estimated with HRXRD around 0.20 % and 0.14 %, respectively.

Direct and reverse SE and spectrophotometry were combined to measure the band gap of the two semiconductors (Ge-VS, $Si_{0.06}Ge_{0.90}Sn_{0.04}$). Figure S2 details the measurement strategy as well as a schematic representation of the SE multilayer model composing each sample. The red arrow refer to a direct (from the top of the sample) or reverse (from the bottom of the sample) SE measurement whereas the blue arrow is either a measurement of reflectance ($R$) when the substrate is single side polished (1SP) or both transmittance ($T$) and $R$ if the substrate is double side polished (DSP). Next, to extract the band gap from these semiconductors, multi-sample analysis (MSA) was used, and an SE model was built and fitted the $R$, $T$, and SE measurement simultaneously whenever possible, as shown in panel (a) and (b) of Figure S2.

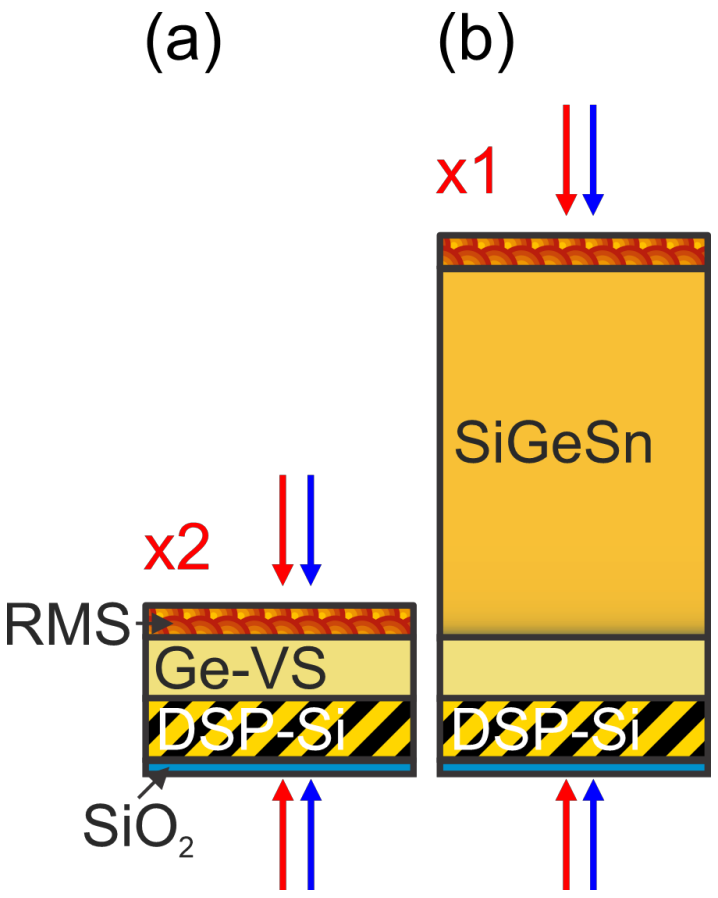

**Figure S2:** Measurement strategy for the (a) Ge-VS, and (b) $Si_{0.06}Ge_{0.90}Sn_{0.04}$ semiconductors. The red arrows indicate the direct (from the top of the sample) or reverse (from the bottom of the sample) SE measurement, whereas the blue arrows are the $R$ and $T$ measurement. For 1SP Si-substrate, only $R$ was measured from the top surface. The shown numbers (x2) indicate the number of grown samples that were characterized with multi-sample analysis.

For reliable fitting of the ellipsometry data of group-IV semiconductor thin films, the following steps were followed:

(1) The multi-sample analysis (MSA) approach was used to model the reverse and direct SE measurement as well as $R$ and $T$ of the 1.5µm thick $Si_{0.06}Ge_{0.9}Sn_{0.04}$ layer grown on a Ge-VS/DSP-Si.

(2) Independent thickness measurement was performed with STEM and APT (Figure 1a-c and Figure 2a-b), surface characterization with AFM (not shown) to quantify the surface roughness layer (RMS), and strain with XRD and RSM maps (Figure 1d-e).

(3) Adequate optical description of Si-substrate is important. To build an accurate optical

model to extract the band gap of the SiGeSn layer, modelling the s- and p-polarized transmission and reflectance spectra at different AOI as well as the direct and reverse ellipsometric data was considered as highlighted previously. However, some difficulties were encountered that hindered the analysis. The issue encountered while fitting the SE parameters was that the agreement of the model with the transmission data was significantly worse than the agreement with the ellipsometry data. This can partly occur due to higher sensitivity of the MSE value to the agreement between experimental and calculated ($\Psi$, $\Delta$) as compared to $T$, and/or due to inadequate description of the optical properties of the substrate. Therefore, we have examined the optical properties of the Si substrate more closely and have performed simultaneous fitting of the transmission and ellipsometry data for the Si substrate. We found that the commonly used Johs-Herzinger model[3] does not adequately describe the substrate transmission and reflection, while good fit can be obtained using a generalized oscillator model to account for the fact that there is a very small absorption in the substrate, induced by the p-type doping of the wafer (B-doped with resistivity of 1-10 Ω.cm), resulting in a significant fit improvement (MSE < 2) (see Figure S3). The blue line in Figure S3 indicates the fitted transmittance calculated with the newly built generalized oscillator optical model.

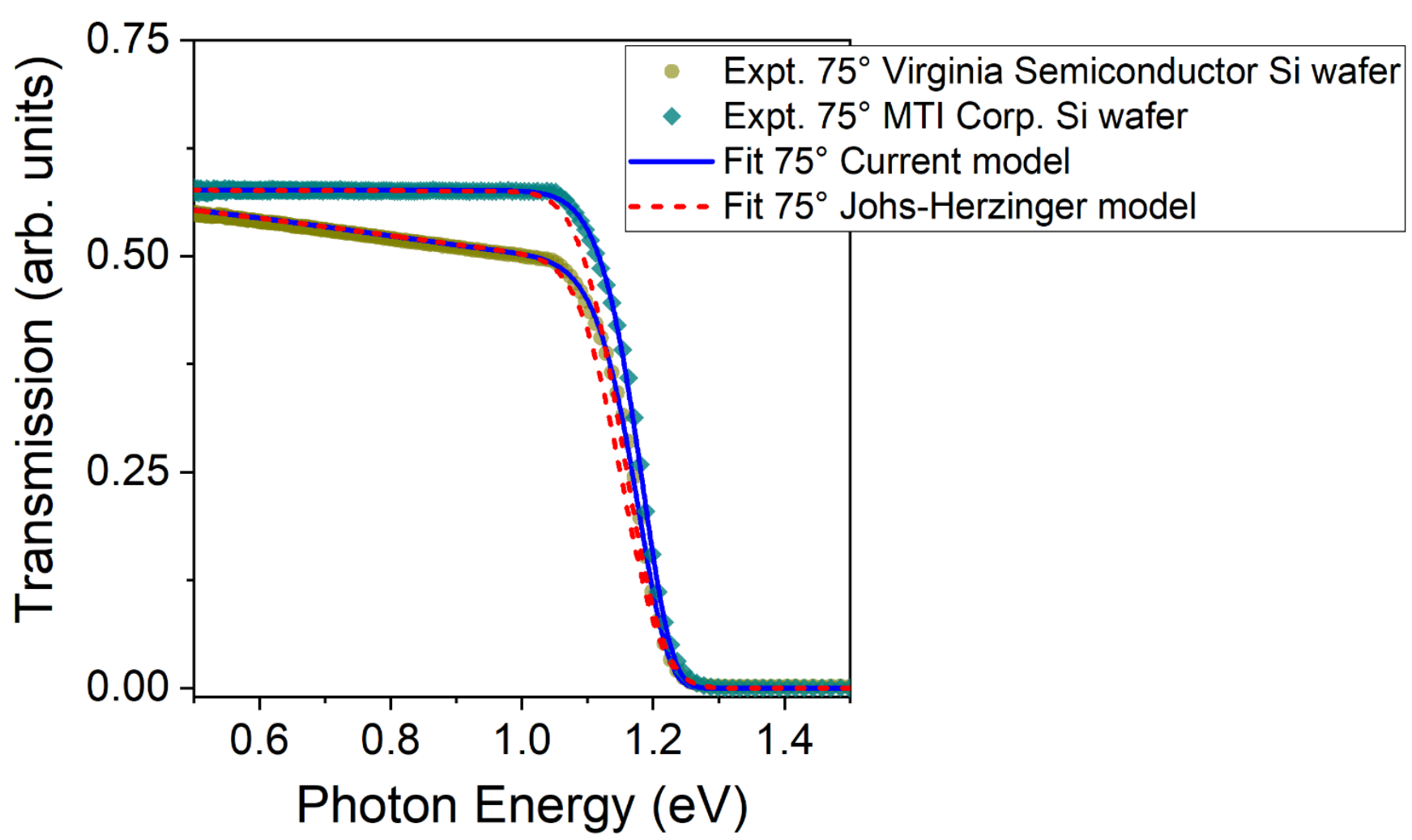

**Figure S3:** Modeled transmittance near Brewster angle 75° of DSP (Virginia Semiconductor and MTI Corp. wafer suppliers) Si substrate with the JH model[3] (dashed red) and the new optical model (blue).

(4) The thickness of the layers in the optical models (Figure S2) is fixed, and the optical

properties ($n$ and $k$) are fitted to the PSEMI model in the entire spectral range. For direct or indirect semiconductors, one of the most sophisticated dispersion models is the parametric semiconductor model (PSEMI) developed by Johs-Herzinger,[4,5] which is based upon the mathematical formalism developed by Garland and Kim.[6] Recently, it has been established that for group-IV tetrahedral semiconductors, the PSEMI model can accurately evaluate the critical points for relaxed and/or strained group-IV alloys [7,8] or thin-film semiconductors. The primary purpose of the PSEMI model is to accurately reproduce the dielectric function of semiconductors with complicated critical point structures ($E_0$, $E_1$, $E_1+\Delta_1$, etc.) in a Kramers-Kronig consistent manner. The result of the built optical model is shown in Figure S4 for the 2 semiconductors where the SE parameter ($\tan\Psi$) is shown at all the spectral range for only 3 AOI (for clarity reasons). The inset in Figure S4 indicate the accuracy of the model (the small MSE also corroborate the quality of the fit) to extract the band gap of the 2 semiconductors as the fitted and experimental data are shown in a small energy range (from 0.5 eV to 1 eV).

(5) Repeat step 4 with and without EMA layer to account for the surface roughness (if roughness is small, improvement with EMA is also small, and parameter uncertainties may be large).

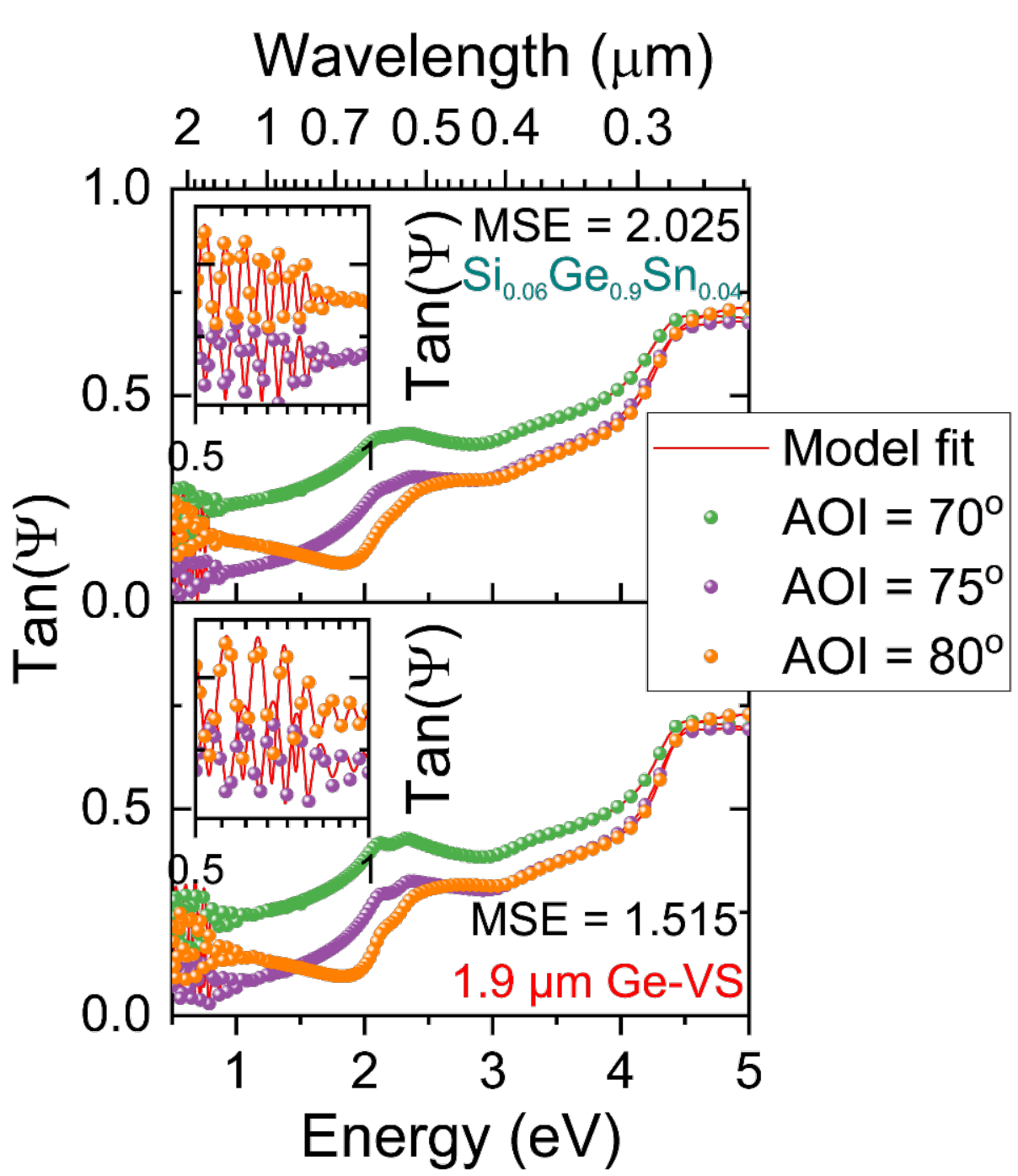

**Figure S4.** Raw (full circles) and fitted (red lines) ellipsometric parameter ($\tan\Psi$) for the different studied semiconductors at three different angles of incidence (70°, 75° and 80°) near the Brewster

angle of Si and Ge. From top panel to bottom: $Si_{0.06}Ge_{0.9}Sn_{0.04}$, and 1.9 μm-thick Ge-VS. The 2 insets correspond to a zoom in the spectral region between 0.5 and 1 eV, as the direct band gap of the 2 semiconductors is hidden within the observed interference fringes.